\documentclass[12pt]{article}
\usepackage{amsmath}
\usepackage{amssymb}

\allowdisplaybreaks

\newtheorem{theorem}{Theorem}
\newenvironment{proof}
  {\noindent \textbf{Proof:}}
  {\qed}

\def\qed{\ifmmode\square\else{\unskip\nobreak\hfil
\penalty50\hskip1em\null\nobreak\hfil$\square$
\parfillskip=0pt\finalhyphendemerits=0\endgraf}\fi}

\newcommand{\Zr}[1]{\ensuremath{\mathbb{Z}/{#1}\mathbb{Z}}}

\newcommand{\Fp}{\ensuremath{\mathbb{F}_p}}
\newcommand{\Fq}{\ensuremath{\mathbb{F}_q}}

\newcommand{\ket}[1]{\ensuremath{|#1\rangle}}

\newcommand{\Tr}{\mathrm{Tr}}

\newcommand{\comment}[1]{}

\begin{document}

\title{Solving Shift Problems and the\\
 Hidden Coset Problem\\
 Using the Fourier Transform}
\author{Lawrence Ip\thanks{Computer Science Division, University of
California, Berkeley, USA. Email: \texttt{lip@eecs.berkeley.edu}.
Supported by NSF Grant CCR-0049092, DARPA Grant F30602-00-2-0601
and DARPA QUIST Grant F30602-01-2-2054. Part of this work was done
while the author was a visitor at the Institute for Quantum
Information at the California Institute of Technology. }}


\maketitle

\begin{abstract}
We give a quantum algorithm for solving a shifted multiplicative
character problem over \Zr{n} and finite fields. We show that the
algorithm can be interpreted as a matrix factorization or as
solving a deconvolution problem and give sufficient conditions for
a shift problem to be solved efficiently by our algorithm. We also
show that combining the shift problem with the hidden subgroup
problem results in a hidden coset problem. This naturally
captures the redundancy in the shift due to the periodic
structure of multiplicative characters over \Zr{n}.
\end{abstract}

\section{Introduction}

The Fourier transform lies at the heart of the solution of all
problems that are known to admit a large quantum speedup over
classical algorithms. This is due to two factors, the quantum
Fourier transform can be performed exponentially faster than its
classical counterpart, and the Fourier transform is particularly
suited to extracting information about periodicities.

Taking a periodic superposition, performing a Fourier transform
and measuring gives us the period. More generally, the Fourier
transform of a superposition over coset states of a subgroup
yields a superposition over the dual or ``perp'' of the subgroup.
This is the basis of the hidden subgroup problem. The most well
known example of a problem and algorithm fitting this framework is
Shor's solution to the factoring problem~\cite{Shor}.

However, information about periodicities is not the only feature
that Fourier transforms can identify. The Fourier transform can
also identify shifts or translations. This suggests that we look
for ways in which the Fourier transform can be used to identify
unknown shifts.

The shift problem may be formulated as follows.

Let $g$ be a complex valued function defined on the group~$G$.
Let $f$ be a shifted version of $g$ where $f(x)=g(x+s)$ for some
$s$ in $G$. Find $s$.

We give an efficient quantum algorithm for solving the shift
problem provided $G$ is abelian and $g$ satisfies certain
conditions. The conditions are satisfied if $G$ is the additive
group of a finite field and $g$ is a multiplicative character. So
our algorithm solves the shifted multiplicative character problem
on a finite field.

If the shift is not uniquely determined, it can be shown that the
set of possible shifts is a coset of a subgroup of the group~$G$.
We then have the \emph{hidden coset problem}.

Let $g$ be a complex valued function defined on the group~$G$.
Let $f$ be a shifted version of $g$ where $f(x)=g(x+s)$ for some
$s$ in $G$. Find all $s'$ satisfying $f(x)=g(x+s')$ for all $x$.

The hidden coset problem combines the features of the hidden
subgroup problem and the shift problem. Our algorithm treats this
problem by first solving a hidden subgroup problem to find $H$ and
then solving a shift problem on the quotient group $G/H$. The
shifted multiplicative character problem over \Zr{n} fits this
framework. The multiplicative characters of \Zr{n} are periodic
with respect to the additive group and so the shift is not
uniquely determined.

Our algorithm solves the shifted Legendre symbol problem and the
shifted quadratic character problem considered by van Dam and
Hallgren~\cite{vanDamHallgren}. The quadratic character is an
example of a multiplicative character and the Legendre symbol is
the quadratic character for \Zr{p}. van Dam and Hallgren also
give an algorithm for the shifted Jacobi symbol problem for \Zr{n}
where $n$ is square free. The Jacobi symbol is also an example of
a multiplicative character. Our treatment solves the problem for
all multiplicative characters of \Zr{n} for all odd $n$.

\section{The Shift Problem}

In this section we state the shift problem, and show how it can be
formulated in terms of matrix multiplication. By factoring the
relevant matrix we obtain an algorithm for solving the shift
problem and also sufficient conditions for the algorithm to be
implemented efficiently on a quantum computer. The analysis
involves Fourier transforms over finite abelian groups. For a
review of characters and Fourier transforms over finite abelian
groups see Appendix~\ref{sect:background}.

The shift problem is the following. Let $g$ be a complex valued
function defined on the group~$G$. Let $f$ be a shifted version
of $g$ where $f(x)=g(x+s)$ for some $s$ in $G$. Find $s$.

We begin by reformulating the shift problem in the following way.
Let $s$ be the unknown shift and $X$ be the matrix with columns
and rows indexed by the elements of $G$ and the matrix element in
row $x$ and column $y$ given by $g(x+y)$. The precise ordering of
the elements of $G$ is irrelevant as long as we are consistent.
That is
\[
X=\left[g(x+y) \right]_{x,y\in G}.
\]
We then compute
\begin{align*}
 X\ket{s}&= \sum_{x\in G} g(x+s) \ket{x}\\
         &= \sum_{x\in G} f(x) \ket{x}.
\end{align*}
Solving the shift problem then reduces to inverting $X$.

We show below that the structure of $X$ allows us to factor $X$
into $X=(F^T)^{-1}DF^{-1}$, where $F$ is the Fourier transform
matrix and $D$ is a diagonal matrix. The diagonal entries of $D$
(up to a scale factor) turn out to be the Fourier transform of $g$
evaluated at the characters of $G$. Provided $g$ satisfies certain
conditions, we can efficiently compute $X^{-1}$ by inverting
$F^T$, $D$ and $F$ in turn to obtain $X^{-1}=F D^{-1} F^T$.

\subsection{Matrix Factorization}

We now show that $X$ is ``diagonalized''\footnote{Strictly
speaking we are not diagonalizing $X$ as that would mean writing
$X=F D F^\dag$ and not $X=FDF^T$.} by the Fourier transform
matrix. Let $F$ be the Fourier transform matrix defined by
\[
F=\left[\psi_y(x) \right]_{x,y\in G},
\]
where $x$ and $y$ index the rows and columns respectively and
$\psi_y$ is a character of $G$ indexed by $y$. Thus each column of
$F$ contains a character of $G$ evaluated at all elements of $G$.

Computing $F^T X F$ we find that
\begin{align*}
F^T X F &=
  \left[ \psi_v(x) \right]_{v,x}
  \left[ g(x+y) \right]_{x,y}
  \left[ \psi_w(y) \right]_{y,w} \\
&=
  \left[ \psi_v(x) \right]_{v,x}
  \left[ \sum_{y\in G} g(x+y)\psi_w(y) \right]_{x,w} \\
&=
  \left[ \psi_v(x) \right]_{v,x}
  \left[  \sum_{y \in G} g(y)\psi_w(y-x) \right]_{x,w} \\
&=
  \left[ \psi_v(x) \right]_{v,x}
  \left[ \overline{\psi_w(x)}
    \sum_{y \in G} g(y)\psi_w(y) \right]_{x,w} \\
&=
  \left[ \psi_v(x) \right]_{v,x}
  \left[ \overline{\psi_w(x)} \hat{g}(\psi_w) \right]_{x,w} \\
&=
  \left[
  \sum_{x\in G} \psi_v(x) \overline{\psi_w(x)}
  \hat{g}(\psi_w) \right]_{v,w} \\
&=
  \left[ |G| \hat{g}(\psi_w) \delta_{vw} \right]_{v,w},
\end{align*}
which is a diagonal matrix. Let $D$ denote this diagonal matrix
and we have $F^TXF = D$.

\subsection{Matrix Inversion}

The algorithm then consists of computing $X^{-1} = F D^{-1} F^T$.
If $X$ is not full rank, then $D$ contains some zeros along the
diagonal. We then calculate the \emph{pseudoinverse} of $X$
(sometimes called the \emph{Moore-Penrose generalized inverse}).
The pseudoinverse of $X$, $X^*$ satisfies
\begin{itemize}
\item $X X^* X = X$,
\item $X^* X X^* = X$,
\item $X X^*$ and $X^* X$ are Hermitian.
\end{itemize}
The pseudoinverse of a matrix always exists and is
unique~\cite{HornJohnson}. The pseudoinverse of a diagonal matrix
is given by inverting the nonzero elements of the diagonal. We
then have $X^*=F D^* F^T$.

The fraction of zeros along the diagonal of $D$ gives the
probability of error, the probability that the algorithm fails to
output $s$.

\subsection{Sufficient Conditions}

\label{sect:suffcond}

To implement the algorithm efficiently we need some conditions on
$g$.
\begin{enumerate}
\item \label{condition:magnitude}
 The magnitude of $g(x)$ is constant for all $x$ such that $g(x)$
 is nonzero.
\item \label{condition:magnitudeFT}
 The Fourier transform of $g$, $\hat{g}(\psi_y)$, has constant
 magnitude for all $y$ for which $\hat{g}(\psi_y)$ is nonzero.
\item \label{condition:efficient}
 $\hat{g}$ can be computed efficiently up to a multiplicative constant
 (on a classical computer).
\end{enumerate}
The above conditions are sufficient for efficient implementation
of the algorithm. The following two parameters determine the
probability of the algorithm outputting the correct answer.
\begin{enumerate}
\item \label{parameter:fraction}
 $\alpha$, the fraction of $x$ in $G$ for which $g(x)$ is nonzero
\item \label{parameter:fractionFT}
 $\beta$, the fraction of $y$ in $G$ for which the
 Fourier transform of $g$, $\hat{g}(\psi_y)$, is nonzero.
\end{enumerate}
The probability of our algorithm succeeding is $\alpha \beta$.

Condition~\ref{condition:magnitude} is needed to be able to create
the superposition with \ket{x} having amplitude $g(x)$ up to a
global constant,
\[
\sum_{x \in G} g(x)\ket{x}.
\]
This can be done efficiently with probability $\alpha$.

Conditions~\ref{condition:magnitudeFT}
and~\ref{condition:efficient} are needed because we will need to
compute $\overline{\hat{g}(\psi_y)}$ to invert $D$.

The parameter $\beta$ is describes the rank of $D$ and thus of
$X$. When we apply $X$ followed by its pseudoinverse, we get a
vector close enough to \ket{s} so that when we measure we obtain
$s$ with probability $\beta$.

Thus the overall probability of success of the algorithm is
$\alpha \beta$.

\subsection{Implementation of the Algorithm}

We now show how to implement the algorithm efficiently on a
quantum computer.

\begin{enumerate}
\item \label{alg:superposition}
Setup a superposition of the all the values of $f$ with the
amplitude of \ket{x} equal to $f(x)$ to obtain
\[
 C\sum_{x \in G} g(x+s)\ket{x}.
\]

\item \label{alg:1stFT}
Compute the Fourier transform to obtain
\[
    C'
    \sum_{y \in G}
    \overline{\psi_y(s)}
    \hat{g}(\psi_y)
    \ket{y}.
\]

\item \label{alg:invertD}
Now compute $\overline{\hat{g}(\psi_y)}$ into the phase to obtain
\[
 C'' \sum_{\substack{y \in G\\\hat{g}(\psi_y)\neq 0}}
 \overline{\psi_y(s)} \ket{y}
\]

\item \label{alg:2ndFT}
Computing the inverse Fourier transform and measuring gives $-s$.
\end{enumerate}
In step~\ref{alg:superposition} we setup a superposition over the
elements of $G$,
\[
\sum_{x\in G} \ket{x},
\]
compute $f(x)$ and measure to see whether $f(x)$ is zero. If so,
then the algorithm fails. If not, we are left with a superposition
over all $x$ such that $f(x)\neq 0$. The algorithm succeeds here
with probability $\alpha$. We next compute $f(x)$ into the
amplitude of \ket{x} (up to a constant factor) .
Condition~\ref{condition:magnitude} ensures that we can do this
by computing the phase of $f(x)$ into the phase of \ket{x}. We
can always approximate this arbitrarily closely by approximating
the phase of $f(x)$ to the nearest $2^n$th root of unity for some
sufficiently large~$n$.

Step~\ref{alg:1stFT} follows from observing that
\begin{align*}
    C'
    \sum_{y \in G}
    \left(
    \sum_{x \in G}
    g(x+s)\psi_y(x)
    \right)
    \ket{y}
&=  C'
    \sum_{y \in G}
    \left(
    \sum_{x \in G}
    g(x)\psi_y(x-s)
    \right)
    \ket{y}\\
&=  C'
    \sum_{y \in G}
    \overline{\psi_y(s)}
    \left(
    \sum_{x \in G}
    g(x)\psi_y(x)
    \right)
    \ket{y}\\
&=  C'
    \sum_{y \in G}
    \overline{\psi_y(s)}
    \hat{g}(\psi_y)
    \ket{y},
\end{align*}
where $C'$ is a constant.

Step~\ref{alg:invertD} can be performed because of
Conditions~\ref{condition:magnitudeFT}
and~\ref{condition:efficient}.

In Step~\ref{alg:2ndFT} we measure and obtain $-s$ with
probability of $\beta$. The reason we get $-s$ instead of $s$ is
that computing the inverse Fourier transform corresponds to
multiplying by $F^\dag$ instead of $F^T$.

Thus the algorithm succeeds in identifying $s$ with probability
$\alpha\beta$ and only requires one query of $f$ and one query of
$\hat{g}$ .

\section{The Hidden Coset Problem}

\label{sect:HiddenCoset}

If $g$ has a ``subgroup'' structure then the shift may not be
unique. We can make this precise in the following way by
formulating a \emph{hidden coset problem} that combines the
features of the shift problem with that of the hidden subgroup
problem. The hidden coset problem is a shift problem where the
shift may not be uniquely defined.

The hidden coset problem is the following. Let $g$ be a complex
valued function defined on the group~$G$. Let $f$ be a shifted
version of $g$ where $f(x)=g(x+s)$ for some $s$ in $G$. Find all
$s'$ satisfying $f(x)=g(x+s')$ for all $x$.

Let $H$ be the largest subgroup of $G$ such that $g$ is constant
on cosets of $H$. Because of the structure of $g$, $s$ is
determined only ``modulo'' $H$. Thus to solve the hidden coset
problem, we need to first identify the hidden subgroup $H$ and
then the shift $s$ modulo $H$.

Assuming that we have already found $H$, to find $s$ modulo $H$,
define $g'$ and $f'$ as complex valued functions on the quotient
group $G/H$ in the natural way so that $g'(x+H) = g(x)$ and
$f'(x+H)=f(x)$ for all $x$ in $G$. Then if $g'$ satisfies the
conditions in Section~\ref{sect:suffcond} when considered over
the group $G/H$ we can apply the algorithm for the shift problem
to find $s$ modulo $H$.

We now show how to find $H$. The standard formulation of the
hidden subgroup problem assumes that $g$ is constant on cosets of
$H$ and that $g$ takes on \emph{distinct values on distinct
cosets}. This can be solved using the ``standard'' algorithm
\begin{enumerate}
\item
Prepare a superposition over all of $G$.
\item
Computing $g$ into a register.
\item
Fourier sampling to obtain a random element of $H^\perp$.
\end{enumerate}

The condition that $g$ takes on distinct values on distinct
cosets of $H$ means that we sample uniformly over the elements of
$H^\perp$. This condition can be relaxed slightly so that the
standard algorithm still works. Boneh and
Lipton~\cite{BonehLipton} and Mosca and Ekert~\cite{MoscaEkert}
give the condition that $g'$ is at most $m$ to 1 and $m$ is less
than the smallest prime factor of $|H|$, the cardinality of $H$.
Hales and Hallgren~\cite{HalesHallgren} give the condition that
at least a polylogarithmic number of values of $g$ need to be
changed to reduce the period of $g$.

However, as we already have restrictions on $g'$ and $\hat{g}'$
(so that we can solve the shift problem) we can give another
condition for the hidden subgroup problem to be solved
efficiently. If $\beta$, the fraction of values of $\hat{g}'$ that
is nonzero satisfies $\beta > 1/p + \text{poly}\log(|H|)$, where
$p$ is the smallest prime factor of $|H|$, then the following
algorithm will find $H$.
\begin{enumerate}
\item
Prepare a superposition over all of $G$.
\item
Computing $g$ into the \emph{phase}.
\item
Fourier sampling to obtain a random element of $H^\perp$.
\end{enumerate}
The difference from the standard algorithm for solving the hidden
subgroup problem is that we compute $g$ into the \emph{phase}.

\section{Shifted Character Problem (Finite Field)}

\label{sect:ShiftFiniteField}

The shifted character problem for the finite field \Fq\ is as
follows.

Given a finite field \Fq\ (where $q=p^m$ for some prime~$p$), a
multiplicative character $\chi$ of \Fq\ and a shifted version of
$\chi$, $f(x)=\chi(x+s)$. Find $s$.

For a review of additive and multiplicative characters in finite
fields see Appendix~\ref{sect:background}.
Appendix~\ref{sect:FTofcharacters} contains a discussion of
Fourier transforms over the additive group of multiplicative
characters.

The shifted multiplicative character problem over a finite field
fits into our general framework. The group $G$ is the additive
group of the finite field. $g=\chi$ is a (non-trivial)
multiplicative character of \Fq. We show that $\chi$ satisfies the
sufficient conditions of Section~\ref{sect:suffcond}.
\begin{enumerate}
\item
 For all nonzero $x$, $\chi(x)$ has unit magnitude.
\item
 $\hat{\chi}(\psi_0)=0$ and as shown in
 Appendix~\ref{sect:FTofcharacters},
 $\hat{\chi}(\psi_{g^m})=
 \overline{\chi (g^m)} \hat{\chi} (\psi_{g^0})$
 and so has magnitude $|\hat{\chi} (\psi_{g^0})|$ which is
 constant.
\item
 $\chi$ and thus $\overline{\chi (g^m)}$ can be
 computed efficiently. So $\hat{\chi}$ can be computed efficiently
 (up to a constant phase).
\end{enumerate}
We next calculate the probability of our algorithm succeeding.
\begin{enumerate}
\item
 $\chi(x)$ is zero only if $x=0$ so $\alpha = 1-1/q$.
\item
 $\hat{\chi}(\psi_y)$ is zero if for $y=0$ so $\beta=1-1/q$.
\end{enumerate}
So our algorithm succeeds with probability $\alpha \beta =
(1-1/q)^2$.

\section{Shifted Character Problem for \Zr{n}}

\label{sect:ShiftZn}

The shifted character problem for \Zr{n} is as follows.

Given a ring \Zr{n} with $n$ odd, a multiplicative character
$\chi$ of \Zr{n} and a shifted version of $\chi$,
$f(x)=\chi(x+s)$. Find $s$.

The shifted character problem over \Zr{n} has the interesting
feature that the solution to the shift is not unique. This is
because multiplicative characters in \Zr{n} have periodicities
with respect to the additive group. See
Appendix~\ref{sect:background} for a discussion of periodicities
of multiplicative characters. Appendix~\ref{sect:FTofcharacters}
contains a discussion of Fourier transforms of multiplicative
characters.

The shifted character problem over \Zr{n} fits into the hidden
coset problem framework described in
Section~\ref{sect:HiddenCoset} so we can apply our algorithm. If
$n=p_1^{m_1} \dots p_k^{m_k}$ we have $\alpha=\beta=(1-1/p_1)
\dots (1-1/p_k)$ and so our algorithm will succeed with
probability $(1-1/p_1)^2 \dots (1-1/p_k)^2$ (after solving the
associated hidden subgroup problem).

\section{Interpretation as Deconvolution}

Our algorithm for solving the shift problem can be thought of
solving a deconvolution problem. To see this, let
$\delta_y(x)=\delta(x-y)$ be the delta function centered at $y$.
Then $f$ is the convolution of $\delta_{-s}$ and $g$, that is
\[
f = \delta_{-s} \star g.
\]
So to recover $s$ or equivalently $\delta_{-s}$, we need to solve
a deconvolution problem.

Taking Fourier transforms and observing that in the Fourier
domain convolution becomes pointwise multiplication we see that
\[
\hat{f} = \hat{\delta}_{-s} \cdot \hat{g},
\]
where $\hat{f}$, $\hat{g}$, $\hat{\delta}_{-s}$ are the Fourier
transforms of $f$, $g$ and $\delta_{-s}$ respectively. We then
have
\[
\delta_{-s} = \mathcal{F}^{-1}  \left( \hat{f} / \hat{g} \right)
\]
where the division is pointwise.

For the division to be performed on efficiently on a quantum
computer would require that the magnitude of $\hat{g}$ be constant
and non-zero. However even if a fraction of the values of
$\hat{g}$ are zero we can still approximate division of $\hat{g}$
by only dividing when $\hat{g}$ is non-zero and doing nothing
otherwise.

Deconvolution is a well studied classical problem and perhaps
this interpretation will enable us to leverage existing
deconvolution techniques to broaden the class of problems
amenable to our approach.

\section{Conclusion}

We have presented a general framework for a class of shift
problems and a set of sufficient conditions for the problems to be
efficiently solved on a quantum computer. However, the sufficient
conditions are fairly restrictive although they include the
shifted multiplicative character problem over finite fields and
rings \Zr{n}.

It would be of interest to investigate what other shift problems
satisfy the sufficient conditions and whether a less restrictive
set of sufficient conditions exists.

\section{Acknowledgements}

I would like to thank Umesh Vazirani for much appreciated advice
and encouragement and Sean Hallgren and Wim van Dam for useful
discussions.

\section{Bibliography}
\bibliographystyle{plain}
\bibliography{writeup}

\begin{thebibliography}{1}

\bibitem{BonehLipton}
Dan Boneh and Richard~J. Lipton.
\newblock Quantum cryptanalysis of hidden linear functions (extended abstract).
\newblock In {\em Advances in Cryptology --- CRYPTO '95}, Lecture Notes in
  Computer Science 963, pages 424--437, 1995.

\bibitem{HalesHallgren}
Lisa Hales and Sean Hallgren.
\newblock An improved quantum fourier transform and appliations.
\newblock In {\em Proceedings 41st Annual Symposium on Foundations of Computer
  Science}, 2001.

\bibitem{HornJohnson}
Roger~A. Horn and Charles~R. Johnson.
\newblock {\em Matrix Analysis}.
\newblock Cambridge, 1985.

\bibitem{LidlNiederreiter}
Rudolf Lidl and Harald Niederreiter.
\newblock {\em Finite Fields}, volume~20 of {\em Encyclopedia of Mathematics
  and Its Applications}.
\newblock Cambridge, second edition, 1997.

\bibitem{MoscaEkert}
Michele Mosca and Artur Ekert.
\newblock The hidden subgroup problem and eigenvalue estimation on a quantum
  computer.
\newblock In {\em Proceedings of the 1st NASA International Conference on
  Quantum Computing and Quantum Communication}, Lecture Notes in Computer
  Science 1509, 1999.

\bibitem{Shor}
Peter~W. Shor.
\newblock Polynomial-time algorithms for prime factorization and discrete
  logarithms on a quantum computer.
\newblock {\em SIAM Journal on Computing}, 26(5):1484--1509, October 1997.

\bibitem{TolimieriAnLu}
Richard Tolimieri, Myoung An, and Chao Lu.
\newblock {\em Algorithms for Discrete Fourier Transform and Convolution}.
\newblock Springer-Verlag, 1989.

\bibitem{vanDamHallgren}
Wim van Dam and Sean Hallgren.
\newblock Efficient quantum algorithms for shifted quadratic character
  problems.
\newblock {\em quant-ph/0011067}, 2000.

\end{thebibliography}
\appendix

\section{Mathematical Background}

\label{sect:background}

We start with some definitions and background. For more details
see the book by Lidl and Niederreiter~\cite{LidlNiederreiter}, and
the book by Tolimieri et al.~\cite{TolimieriAnLu}.

\subsection{Characters of a Group}

A character~$\chi$ of a finite abelian group $G$ is a homomorphism
from a group~$G$ to $(\mathbb{C},\times)$, the group of complex
numbers with multiplication. That is
\[
\chi(g_1 g_2) = \chi(g_1) \chi(g_2)
\]
for all $g_1, g_2 \in G$. If $G$ is cyclic with generator~$g$, the
characters are
\[
\chi_k(g^l) = \exp(2 \pi i kl/|G|)
\]
for $k=0,\dots,|G|-1$.

The characters of $G$ form a group $\hat{G}$, known as the dual
group, with multiplication defined as
\[
(\chi_1 \chi_2)(g) = \chi_1(g) \chi_2(g)
\]
for all $\chi_1, \chi_2 \in \hat{G}$, $g\in G$. $\hat{G}$ is
isomorphic to $G$. In particular, $|\hat{G}|=|G|$, that is the
number of characters of $G$ is the same as the cardinality of $G$.

\subsection{Fourier Transform over a Group}

Given a complex valued function~$f$ over $G$, the Fourier
transform of $f$ over the group $G$ is given by
\[
\hat{f}(\chi) = \sum_{g \in G} f(g) \chi(g)
\]
for all $\chi \in \hat{G}$.

\subsection{Characters of a Field}

In a field \Fq\ (where $q=p^m$ for some prime~$p$) we have two
operations, multiplication and addition, with corresponding
groups. Thus we can define two different groups of characters.

Characters of the multiplicative group $\Fq^*$ of \Fq\ are called
\emph{multiplicative characters} of \Fq. Since $\Fq^*$ is cyclic
with order $q-1$, its characters $\chi_0,\dots\chi_{q-2}$ can be
explicitly represented as
\[
\chi_k(g^l) = \exp\left(\frac{2 \pi i kl}{q-1} \right)
\]
for all $l=0,1,\dots,q-2$, where $g$ is a generator of $\Fq^*$. It
is often convenient to extend the definition of $\chi_k$ to
include 0 by defining $\chi_k(0)=0$. The quadratic character
referred to in van Dam and Hallgren~\cite{vanDamHallgren} is
$\chi_{\frac{q-1}{2}}$.

Characters of the additive group of \Fq\ are called \emph{additive
characters} of \Fq. The additive characters $\psi_a$ (for all $a
\in \Fq$) have the form
\[
\psi_a(c)=\exp\left(\frac{2 \pi i \Tr(ac)}{p}\right)
\]
for all $c \in \Fq$, where
\[
\Tr(x) = \sum_{k=0}^{m-1} x^{p^k}
\]
is the trace function from \Fq\ to \Fp\ and \Fp\ is identified
with \Zr{p} for the purposes of evaluating the exponential.
$\psi_1$ is the canonical additive character of \Fq. We have that
$\psi_a(c) = \psi_1(ac)$.

\subsection{Characters of \Zr{n}}

\label{CharactersZn}

Similarly we can define additive and multiplicative characters in
\Zr{p^m} where $p$ is an odd\footnote{We only consider odd $p$
because $(\Zr{2^m})^*$ is not cyclic for $m \geq 3$. In fact,
$(\Zr{2^m})^*$ is the product of two cyclic
groups~\cite{TolimieriAnLu}.} prime. \Zr{p^m} with addition is a
cyclic group so we can define additive characters $\psi_0,
\dots,\psi_{p^m-1}$
\[
\psi_k(x) = \exp \left( \frac{2 \pi i kx}{p^m} \right)
\]
for all $x \in \Zr{p^m}$.

$(\Zr{p^m})^*$ is a cyclic group, so we can define multiplicative
characters $\chi_0,\dots,\chi_{(p-1)p^{m-1}-1}$, where
\[
\chi_k(g^l) = \exp \left( \frac{2 \pi i kl}{(p-1)p^{m-1}} \right)
\]
for all $l=0,1,\dots,(p-1)p^{m-1}-1$ and $g$ is a generator of
$(\Zr{p^m})^*$. We can extend the definition of $\chi_k$ to
include all of \Zr{p^m} by defining $\chi_k(x)=0$ if $x$ is a
multiple of $p$.

\Zr{n} with addition is a cyclic group so we can define additive
characters $\psi_0, \dots,\psi_{n-1}$
\[
\psi_k(x) = \exp \left( \frac{2 \pi i kx}{n} \right)
\]
for all $x \in \Zr{n}$.

If $n$ is odd, we can define multiplicative characters over \Zr{n}
by observing that if $n=p_1^{m_1}\dots p_k^{m_k}$ then
\[
\Zr{n} \cong \Zr{p_1^{m_1}} \oplus \dots \oplus \Zr{p_k^{m_k}}.
\]
Let $\phi$ be an isomorphism from \Zr{n} to $\Zr{p_1^{m_1}} \oplus
\dots \oplus \Zr{p_k^{m_k}}$ with $\phi(x) =
(\phi_1(x),\dots,\phi_k(x))$ given by $\phi_j(x)= x \bmod
p_j^{m^j}$. Define the multiplicative characters as the product of
the multiplicative characters in the corresponding \Zr{p_j^{m_j}}.
That is
\[
 \chi_{l_1,\dots,l_k}(x)
 = \prod_{r=1}^k \chi_{l_r}^{(r)}(\phi_r(x)),
\]
where $\chi_{l_r}^{(r)}$ is a multiplicative character of
\Zr{p_r^{m_r}}. Noting that all the $\chi_{l_1,\dots,l_k}$ are
distinct for appropriate ranges of $l_1,\dots,l_k$ and that there
are exactly the right number of them we see that we have all the
multiplicative characters.

\subsection{Periodicities of Multiplicative Characters of
\Zr{p^m}}

The multiplicative characters of \Zr{p^m} have a particular
\emph{additive} periodic structure. Every multiplicative character
has a period that is a power of $p$. That is, for every~$k$, there
is a $j$ such that
\[
\chi_k(x+p^j) = \chi_k(x)
\]
for all $x \in \Zr{p^m}$.

In fact we have the following theorem~\cite{TolimieriAnLu} that
explicitly gives the period.
\begin{theorem}
Let~$g$ be a generator of the group of units $(\Zr{p^m})^*$ and
let $\chi_k$ be a multiplicative character of \Zr{p^m} defined
\[
\chi_k(x)=
   \begin{cases}
      0&    \text{if $p|x$},\\
      \exp\left(\frac{2\pi i kl}{(p-1)p^{m-1}}\right)
        &  \text{if $x=g^l$}.
   \end{cases}
\]
Let~$j$ be such that $\gcd(p^m,k)=p^{m-j}$. Then $p^j$ is the
additive period of $\chi_k$, the smallest $T$ such that
$\chi_k(x+T)=\chi_k(x)$ for all $x \in \Zr{p^m}$.
\end{theorem}
\begin{proof}
Suppose $\gcd(p^m,k)=p^{m-j}$ and that the period of $\chi_k$ is
$T$. We first show that $T|p^j$ and then that $T=p^j$.

If $p|x$ then $p|(x+p^j)$ and so $\chi_k(x+p^j)=0=\chi_k(x)$. If
$x=g^l$ then because $p^{m-j} | k$, the value of $\chi_k(g^l)$ is
determined by the value of $l \bmod (p-1)p^{j-1}$, which is in
turn determined by the value of $g^l \bmod p^j$. But $g^l+p^j
\equiv g^l \mod p^j$ and so $\chi_k(x+p^j)=\chi_k(x)$. This shows
that $T|p^j$.

Since $T|p^j$, if $T\neq p^j$, we must have $T=p^{j'}$ for some
$j'<j$. Then $\chi_k(1)=\chi_k(1+rp^{j'})$ for all
$r=0,\dots,p^{m-j'}-1$. Now
\[
 \{1+rp^{j'} : r=0,\dots,p^{m-j'}-1\}
 =
 \{g^{(p-1)p^{j'-1}l'} : l'=0,\dots,p^{m-j'}-1\}
\]
since all the $g^{(p-1)p^{j'-1}l'}$ are distinct and
$g^{(p-1)p^{j'-1}l'}\equiv 1 \mod p^{j'}$. But
$\chi_k(g^{(p-1)p^{j'-1}l'})=1$ for all $l'$ only if $k (p-1)
p^{j'-1}$ is a multiple of $(p-1)p^{m-1}$ which implies that
$p^{m-j'} | k$. This is a contradiction since $p^{m-j}$ was the
largest power of $p$ dividing $k$. Thus $T=p^j$.
\end{proof}

\subsection{Periodicities of Multiplicative Characters of \Zr{n}}

As we saw in Appendix~\ref{CharactersZn}, a multiplicative
character $\chi_{l_1,\dots,l_k}$ of \Zr{n} can be decomposed into
a product of multiplicative characters
$\chi_{l_1}^{(1)}\dots\chi_{l_k}^{(k)}$ of $\Zr{p_1^{m_1}}, \dots,
\Zr{p_k^{m_k}}$. The period of $\chi_{l_1,\dots,l_k}$ is then the
product of the periods of $\chi_{l_1}^{(1)}\dots\chi_{l_k}^{(k)}$.

\section{Fourier Transforms of Multiplicative Characters}

\label{sect:FTofcharacters}

\subsection{Finite Field Case}

Consider the natural representation of $\mathbb{F}_q^*$, the
multiplicative group. It is well known that this is cyclic. If $g$
is a generator then the elements can be represented as $\{1, g,
g^2, g^3, \cdots, g^{q-2} \}$. The multiplicative characters are
then exponentials in $k$
\[
\chi_l(g^k) = \exp \left( \frac{2 \pi i k l}{q-1} \right).
\]
The non-trivial additive characters can be written
\[
\psi_{g^m}(g^k) = \exp \left( \frac{2 \pi i \text{Tr}(g^m g^k)}{p}
\right).
\]
Note that these are all translates of the canonical additive
character
\[
\psi_{g^0}(g^k) = \exp \left( \frac{2 \pi i \text{Tr}(g^k)}{p}
\right).
\]

When we take the Fourier transform of a multiplicative character
over the additive group we are expressing an exponential in terms
of a basis where the basis functions are translates of the
canonical additive character. To compute the change of basis we
compute the inner product of the exponential with all the
translates of the canonical additive character. This is the same
as the inner product of the canonical additive characters with the
exponential translated in the opposite direction. But a translated
exponential is just the original exponential with a phase shift
that is given by the exponential of the size of the translation.
Thus
\begin{align*}
 \hat{\chi}_l(\psi_{g^m})
 &=
 \sum_{k=0}^{q-2} \psi_{g^m}(g^k) \chi_l(g^k)\\
 &=
 \sum_{k=0}^{q-2} \exp \left( \frac{2 \pi i \text{Tr}(g^m
 g^k)}{p} \right)
 \exp \left( \frac{2 \pi i k l}{q-1} \right)\\
 &=
 \sum_{k=0}^{q-2} \exp \left( \frac{2 \pi i (k-m) l}{q-1} \right)
 \exp \left( \frac{2 \pi i \text{Tr}(g^k)}{p} \right)\\
 &=
 \exp \left( -\frac{2 \pi i m l}{q-1} \right)
 \hat{\chi}_l(\psi_{g^0})\\
 &=
 \overline{\chi_l(g^m)} \hat{\chi}_l(\psi_{g^0}).
\end{align*}

\subsection{\Zr{p^m} Case}

If a multiplicative character $\chi_l$ of \Zr{p^m} has no
periodicity, then $p\nmid l$. An extension of the argument used
for the finite field case shows that
\[
 \hat{\chi}_l(\psi_{y}) =
 \overline{\chi_l(y)} \hat{\chi}_l(\psi_{1}).
\]
See Tolimieri et al. for details~\cite{TolimieriAnLu}. If
$\gcd(p^m,l)=p^{m-j}$, $\chi_l$ has period $p^l$ and the previous
argument does not work because $p^m$, the size of the additive
group, and $l$ are not relatively prime. However, by projecting
\Zr{p^m} onto \Zr{p^j} by sending $x$ to $x\bmod p^j$ we
transform $\chi_l$ to a multiplicative character of \Zr{p^j} with
no periodicity. Thus the Fourier transform of $\chi_l$ over
\Zr{p^m} must be
\[
 \hat{\chi}_l(\psi_{y}) =
 \begin{cases}
 K \overline{\chi_l(y/p^{m-j})}
 & \text{if $p^{m-j}|y$},\\
 0 & \text{if $p^{m-j}\nmid y$},
 \end{cases}
\]
for some constant $K$.

\subsection{\Zr{n} Case}

Given a multiplicative character of \Zr{n}, $\chi_{l_1,\dots,l_k}$
the vector with component $x$ equal to $\chi_{l_1,\dots,l_k}(x)$
is equal to the tensor product of the corresponding vectors of the
$\chi_{l_1}^{(1)}, \dots, \chi_{l_k}^{(k)}$.

The Fourier transform of $\chi_{l_1,\dots,l_k}$ will then be the
tensor product of the Fourier transforms of the
$\chi_{l_1}^{(1)}, \dots, \chi_{l_k}^{(k)}$. So if the period of
$\chi_{l_1,\dots,l_k}$ is $T$, the Fourier transform of
$\chi_{l_1,\dots,l_k}$ is
\[
 \hat{\chi}_{l_1,\dots,l_k}(\psi_y) =
 \begin{cases}
 K \overline{\chi_{l_1,\dots,l_k}(y/T)}
 & \text{if $T|y$},\\
 0 & \text{if $T\nmid y$},
 \end{cases}
\]
for some constant $K$.

\end{document}